\documentclass[aps,prl,twocolumn,showpacs]{revtex4-1}
\usepackage{graphicx}
\usepackage{amsmath}
\usepackage{amsfonts}
\usepackage{amssymb}
\usepackage{bm}

\def\bea{\begin{eqnarray}}
\def\eea{\end{eqnarray}}
\def\be{\begin{equation}}
\def\ee{\end{equation}}
\def\bpm{\begin{pmatrix}}
\def\epm{\end{pmatrix}}

\def\nn{\nonumber}
\def\Im{\mathop{\rm Im}}

\newcommand{\corr}[1]{\langle #1\rangle}

\newcommand{\p}{\partial}

\newcommand{\D}{\Delta}

\newcommand{\bk}{{\bm k}}

\newcommand{\bj}{{\bm j}}
\newcommand{\bi}{{\bm i}}
\newcommand{\bS}{{\vec{S}}}

\newcommand{\bde}{{\bm e}}

\newcommand{\bn}{{\bm d}}
\newcommand{\n}{{ d}}

\newcommand{\I}{{\rm i}}
\newcommand{\dt}{\text{o}}
\newcommand{\ds}{\text{e}}

\begin{document}

\title{Exotic $S=1$ spin liquid state with fermionic excitations on triangular lattice}
\author{Maksym Serbyn, T. Senthil,  and Patrick A. Lee}
\affiliation{Department of Physics, Massachusetts Institute of
Technology, Cambridge, Massachusetts 02139}
\date{\today}

\begin{abstract}
Motivated by recent experiments on the material Ba$_3$NiSb$_2$O$_9$ we consider a spin-one quantum antiferromagnet on a triangular lattice with the Heisenberg bilinear and biquadratic exchange interactions and a single-ion anisotropy. Using a fermionic ``triplon'' representation for spins, we study the phase diagram within mean field theory. In addition to a fully gapped spin-liquid ground state, we find a state where one gapless triplon mode with Fermi surface coexists with $d + id$ topological pairing of the other triplons. Despite the existence of a Fermi surface, this ground state has fully gapped bulk spin  excitations. Such a state has linear in temperature specific heat and constant in plane spin susceptibility, with an unusually high Wilson ratio.

\end{abstract}


\maketitle

Spin liquid (SL) is a long sought exotic state of matter proposed by Anderson~\cite{Anderson1,*Anderson2}, where long range magnetic order is destroyed by quantum fluctuations at zero temperature. A number of  materials have been discovered which are promising candidates for two-dimensional $S=1/2$ SL state~\cite{PAL-Science}. More recently, possible SL materials with $S=1$  have been discussed.  One example is the insulating spin-1 quantum magnet on a triangular lattice, NiGa$_2$S$_4$, reported by Nakatsuji \emph{et al}~\cite{Nakatsuji}. This material motivated a number of theoretical papers proposing different microscopic realizations of $S=1$ SL~\cite{Senthil-NiGaS,Tsunetsugu,Ng-short,*Ng-long,GroverSenthil}. Recently high pressure synthesis of the two-dimensional triangular magnet Ba$_3$NiSb$_2$O$_9$~\cite{exp} has produced two new phases which possibly realize two and three-dimensional $S=1$ SL. In particular the 6H-B phase, described as a triangular lattice of  Ni$^{2+}$ ions, shows no magnetic ordering down to $T=350$~mK along with linear in temperature specific heat~(with unusually high coefficient) and constant spin susceptibility. The metal-like behavior of specific heat and spin susceptibility observed in the insulating 6H-A phase suggest the presence of quasiparticle excitations with a Fermi surface.

Motivated by this newly discovered material, in the present Letter we propose a new candidate SL ground state with exotic physical properties. Our model system consists of quantum $S=1$ spins forming a triangular lattice. For simplicity, we consider only nearest neighbor interactions. The general form of Hamiltonian can be written as
\be \label{HHeis}
  H
  =
  \sum_{\corr{\bi\bj}}
  [J \bS_\bi\cdot\bS_\bj
  +
  K (\bS_\bi\cdot\bS_\bj)^2]
  +D \sum_\bi (S_\bi^z)^2,
\ee
\begin{figure}
\includegraphics[width=0.9\linewidth]{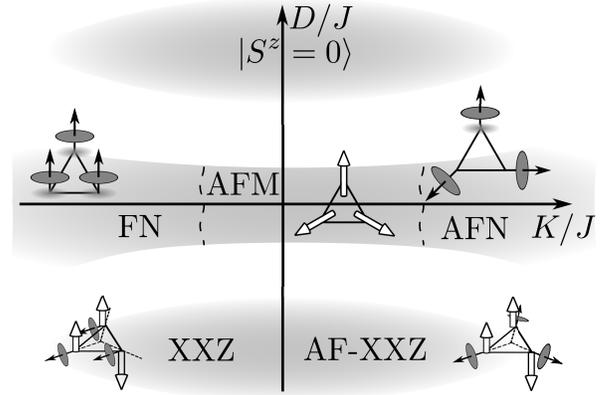}
\caption{\label{Fig:GS} Schematic representation of the ground state in different limits of the Hamiltonian~(\ref{HHeis}). White arrows represent average spin; arrows with discs indicate the director of the nematic order parameter. Details are discussed in the text. }
\end{figure}%
where we included Heisenberg exchange interaction with coupling $J>0$ and biquadratic exchange with coupling~$K$. In addition we allow easy-plane or easy-axis type of anisotropy controlled by the parameter~$D$, but we neglect this anisotropy in the couplings~$J$ and~$K$ since it is presumably small for transition metals. The Hamiltonian~(\ref{HHeis}) has been considered in the literature in limits when the anisotropy is either zero or dominates over other couplings, or there are longer range competing exchange couplings.  Fig.~\ref{Fig:GS} summarizes known results for the ground state (GS) phase diagram in a schematic way. There are three different phases on the line of zero anisotropy $D=0$~\cite{Papa,Lauchli,Toth,Peng}: in the range $K=-0.4 J \ldots J$ GS is $120^\circ$-degree antiferromagnet (AFM). For larger negative $K$ system favors collinear ferro-nematic (FN) order, \emph{i.e.} nematic order that does not break lattice translational symmetry. In this state the average spin vanishes $\corr{\bS}=0$, but full spin rotation symmetry is broken down  to rotations around an axis specified by the director vector $\bn$~(see Refs.~\cite{Lauchli,Toth} and discussion below). For positive $K>J$ the ground state is described by aniferro-nematic (AFN) order. In this state director vectors $\bn_\bi$ on three different sublattices are orthogonal to each other~(see Fig.~\ref{Fig:GS}), thus breaking lattice translation symmetry.
In the extreme case of easy-plane anisotropy~($D\gg J, |K|$), the GS is a trivial product of states of $|S^z=0\rangle$ on all sites, corresponding to the trivial single-site FN order. For large but negative~$D$, implying extreme easy axis anisotropy,  only two states with $S^z=\pm 1$ on each site survive. This system can be described by a spin-$1/2$ XXZ model with all exchange couplings being antiferromagnetic if $2J>K>0$ or with $J^z$ being frustrating and $J^\perp$ ferromagnetic if $K<0$. In both cases there is spin density wave ordering of the $z$-component of the spin in the GS, supplemented by planar AFN order in former and collinear nematic order in the latter case~\cite{DamleSenthil}.

Physically for Ba$_3$NiSb$_2$O$_9$ we may expect the exchange coupling $J$ to be the largest with $J>|K|,|D|$. Both signs of $D$ seems plausible. Likewise it is not known what sign of the biquadratic exchange $K$ is realized, even though negative $K$ can be obtained from large $U$ expansion of certain multi-orbital Hubbard model or from coupling to phonons. Therefore in what follows we study the phase diagram of Hamiltonian~(\ref{HHeis}) for both signs of $D$ and $K$ but will assume $|D|,|K|<J$. Except of very small $|D|$, this is outside of the regions of known GS's shown in Fig.~\ref{Fig:GS}. In order to
to get access to the (RVB-like) state with fermion excitations, we use the fermion representation of the spin~\cite{Ng-long}. After this we study resulting phase diagram in the mean field  approximation.

\emph{Fermion representation.}
The spin operator is conveniently represented via a set of three operators called triplons which are labeled by index $\alpha=x,y,z$. In earlier papers~\cite{Papa,GroverSenthil}
these operators were bosons, but here we use fermions~\cite{Ng-long} written as a vector $\vec{f}_\bi=(f_{\bi x},f_{\bi y},f_{\bi z})^T$,
\be
  \vec{S}_\bi
  =
  - \I \vec{f}_{\bi}^\dagger \times \vec{f}_{\bi}
   ,
   \qquad
   \vec{f}_\bi^\dagger \cdot \vec{f}_\bi
  =
  1.
  \label{constraint}
\ee
In terms of $S^z$ eigenstates, we used the following basis to represent the states of $S=1$, $|x\rangle=\I (|1\rangle-|-1\rangle)/\sqrt{2}$, $|y\rangle=(|1\rangle+|-1\rangle)/\sqrt{2}$, $|z\rangle=-\I |0\rangle$, since it facilitates the handling of the biquadratic term in the Hamiltonian. Eq.~(\ref{constraint}) also imposes a constraint of single occupation in order to exclude unphysical states from the Hilbert space. In the mean field theory this constraint will be relaxed to hold only on average. There are two possible choices of constraint for spin-one system: the particle representation that we used above  and the hole representation $\vec{f}_\bi^\dagger \cdot \vec{f}_\bi=2$. In contrast to the case of $S=1/2$, these are not equivalent. Nevertheless, they can be mapped into each other by particle-hole transformation plus a change of the sign of hopping. Therefore we consider only particle representation but do not restrict hopping to be positive to include the hole representation~\footnote{This is different from Ref.~\cite{Ng-short}, where authors use combination of particle and hole constraints in order to preserve particle-hole symmetry. Our treatment violates particle-hole symmetry from the very beginning.}.

The chosen spin representation has $U(1)$ redundancy remaining~\cite{Ng-long,Note1}: one can multiply $\vec{f}_\bi$ by a phase factor leaving the spin intact. In addition, in the absence of $D$ there is a spin rotation symmetry, realized by the simultaneous rotation of the vectors $\vec{f}_\bi$ and $(\vec{f}_\bi^\dagger)^T$. Non-zero anisotropy $D$ breaks full spin rotation symmetry to rotation symmetry in $xy$-plane supplemented by the reflection of spin along $z$-axis.

The bilinear term is expressed via fermions as
$
  \bS_\bi\cdot \bS_\bj
  =
  (\vec{f}^\dagger_\bi\cdot\vec{f}^\dagger_\bj)
  (\vec{f}_\bi\cdot\vec{f}_\bj)
  +
  \vec{f}^\dagger_\bi (\vec{f}_\bi\cdot\vec{f}^\dagger_\bj)\vec{f}_\bj.
$
Using the constraint $\vec{f}_\bi^\dagger \cdot\vec{f}_\bi=1$, the biquadratic term also can be expressed as a product of four fermion operators~\cite{Papa},
$
  (\bS_\bi\cdot \bS_\bj)^2
  =
  1
  -
  (\vec{f}^\dagger_\bi\cdot\vec{f}^\dagger_\bj)
  (\vec{f}_\bi\cdot\vec{f}_\bj).
$
Adding a Lagrange multiplier to enforce the single occupancy constraint~(\ref{constraint}) on average, we have
\begin{multline} \label{Hf}
  H
  =
  \sum_{\corr{\bi\bj}}
  [
   J \vec{f}^\dagger_\bi (\vec{f}_\bi\cdot\vec{f}^\dagger_\bj)\vec{f}_\bj
   +
   (J-K)(\vec{f}^\dagger_\bi\cdot\vec{f}^\dagger_\bj)(\vec{f}_\bi\cdot\vec{f}_\bj)
   +
   K
   ]
  \\
  + \sum_\bi [\mu(1-\vec{f}_\bi^\dagger \cdot\vec{f}_\bi)+D(1-f^\dagger_{\bi z}f_{\bi z})],
\end{multline}

\emph{Mean field results.}
Having expressed the Hamiltonian via fermion operators we study the  mean field phase diagram of our model.  To unambiguously decouple quartic fermion terms, we use the Feynman variational principle~\cite{Feynman,LeeF} which is equivalent to the trial wave functions approach.
We define an action based on the Hamiltonian~(\ref{Hf}),
$
  S
  =
  \int_0^\beta d\tau\,
  \big[\sum_\bi f^\dagger_{\bi\alpha} (\p_\tau-\mu)f_{\bi\alpha}+ H \big],
$
as well as the trial quadratic action, $\tilde S$, with $H$ replaced by $\tilde H$,
\be \label{tildeH}
  \tilde H
  =
  \sum_{\corr{\bi\bj}}
  [\vec{f}^\dagger_\bi T_{\bi\bj} \vec{f}_\bj+
  \vec{f}^\dagger_\bi A_{\bi\bj} \vec{f}^\dagger_\bj
  +
  \text{H.c.}]
  +
  \sum_\bi
  \vec{f}^\dagger_\bi t_{\bi} \vec{f}_\bi.
\ee
The mean field parameters $T_{\bi\bj}$, $A_{\bi\bj}$, and $t_\bi$ are determined from the stationary points of the functional $\Psi[\tilde S] = \corr{S-\tilde S}_{\tilde S} - \log \tilde Z$,
\begin{subequations} \nn \label{TAMF}
\bea \nn\label{TMF}
  T_{\bi\bj}^{\alpha\beta}
  &=&
  - J\, \delta_{\alpha\beta} \corr{f^\dagger_{\bj\kappa} f_{\bi\kappa}}
  +(J-K) \corr{f^\dagger_{\bj\alpha} f_{\bi\beta}}
  ,
  \\ \nn\label{AMF}
   A_{\bi\bj}^{\alpha\beta}
  &=&
  - J \corr{f_{\bi\beta} f_{\bj\alpha}}
  + (J-K)\,\delta_{\alpha\beta} \corr{f_{\bi\kappa} f_{\bj\kappa}}
  ,
  \\ \nn\label{tMF}
   t_\bi^{\alpha\beta}
   &=&
  \sum_{\corr{\bi\bj}}
  [J\corr{f^\dagger_{\bj\beta}f_{\bj\alpha}}
  -
  (J-K)\corr{f^\dagger_{\bj\alpha}f_{\bj\beta}}
  ]
  -D\delta_{\alpha \beta}\delta_{\alpha z}
  .
\eea
\end{subequations}
For $T=0$, we get  the estimate for the ground state energy, $E_\text{g.s.}\leq\tilde E_\text{g.s.} =\corr{H}_{\tilde S}$, where
\begin{multline} \label{Egsp}
  \tilde E_\text{g.s.}
  =
  \sum_{\corr{\bi\bj}}
  [
  T^{\alpha\beta}_{\bi\bj}\corr{f^\dagger_{\bi\alpha}f_{\bj\beta}}
  +
  A^{\alpha\beta}_{\bi\bj}\corr{f^\dagger_{\bi\alpha}f^\dagger_{\bj\beta}}]
  \\
  +\frac12\sum_\bi
  [t_\bi^{\alpha\beta} \corr{f^\dagger_{\bi\alpha}f_{\bi\beta}}
  -D\corr{f^\dagger_{\bi z}f_{\bi z}}
  +6K+2D
  ].
\end{multline}

We search for self-consistent solutions to the mean field equations that do not break any additional symmetries other than $\cal T$-reversal. When the full spin rotation symmetry is present, the only possible pairing order parameter is $\D_\dt \sim \corr{\vec{f}_\bi \cdot \vec{f}_\bj}$. Such pairing preserves full rotational symmetry in spin space, the resulting state being a spin singlet. We call this pairing in odd channel, since it is possible only with odd orbital momentum, \emph{i.e.} $p$, $f$-wave pairing. Since in Hamiltonian~(\ref{HHeis}), only \emph{in-plane} rotational symmetry is present for $D\neq0$, the pairing in even channel with order parameter $\D_\ds \sim \corr{(\vec{f}_\bi \times \vec{f}_\bj)_z}=\corr{f_{\bi x} f_{\bj y}-f_{\bi y}f_{\bj x}}$ is allowed. However, the presence of two order parameters simultaneously violates the symmetry with respect to rotations of $\pi$ around the $x$ or $y$ axis.

Both aforementioned types of pairing were considered by Liu \emph{et.al.}~\cite{Ng-long} in a similar system, however without anisotropy but with competing third nearest neighbor $J$. Their treatment of biquadratic exchange also differs from ours. The result of~\cite{Ng-long} was that pairing in odd channel always wins. Below, after establishing the mean field equations for each type of pairing,  we identify the region in phase space where even-channel pairing has lower energy than  odd-channel pairing.

\emph{Pairing in odd channel.} We introduce the mean field parameters $\chi^{\alpha}$, $n^{\alpha}$, and $\Delta^{\alpha}_\dt$, $\alpha=x,y,z$ defined as
\be \label{chiD}
  \chi^{\alpha}
  =
  \corr{f^\dagger_{\bi \alpha} f_{\bi+\bde_1 \alpha}},
  \quad
  n^\alpha
  =
  \corr{f^\dagger_{\bi \alpha} f_{\bi \alpha}},
  \quad
  \D^\alpha_\dt
  =
  \corr{f_{\bi \alpha} f_{\bi+\bde_1 \alpha}}.
\ee
The vectors  $\bde_1=(1,0)$, $\bde_2=(1/2,\sqrt{3}/2)$, and $\bde_3=\bde_2-\bde_1$ specify link orientation. The hopping is the same on all links, whereas the pairings for the remaining two orientations are $\corr{f_{\bi \alpha} f_{\bi+\bde_2 \alpha}}=\D^\alpha_\dt e^{\I\pi l \over3}$,  $\corr{f_{\bi \alpha} f_{\bi+\bde_3 \alpha}}=\D^\alpha_\dt e^{2\I\pi l \over3}$, where the pair angular momentum $l=1,2,3$ for $p+\I p$, $d+\I d$, and $f$-wave pairing respectively. Spin rotation symmetry in the $xy$-plane requires $\chi^x=\chi^y$, $n^x=n^y$, $\D^x_\dt=\D^y_\dt$. The Hamiltonian in momentum space (modulus non-essential constant terms) can be rewritten as
\be \label{Htriplet}
  \tilde H
  =
  \sum_{\bk,\alpha}
  \chi^\alpha_\bk f^\dagger_{\bk \alpha}f_{\bk \alpha}
  +
  \D_\bk^\alpha  f^\dagger_{\bk \alpha}f^\dagger_{-\bk \alpha}
  +
  \D_\bk^{\alpha *}  f_{-\bk \alpha}f_{\bk \alpha},
\ee
with mean field parameters
\bea
\label{ChiMF}
   \chi^\alpha_\bk
   &=&
   2\gamma(\bk)[(J-K)\chi^\alpha
   -J(\chi^x+\chi^y+\chi^z)]
   \\\nn
   &&+6K n^\alpha -\mu-\delta_{\alpha, z} D,
   \\ \label{Dtriplet}
   \D^\alpha_\bk
  &=&
  \psi(\bk)
  [(J-K) (\D^x_\dt+\D^y_\dt+\D^z_\dt)-J \D^\alpha_\dt].
\eea
The function $\gamma(\bk)$ is a sum over nearest neighbors,
$
  \gamma(\bk)
  =
  \cos \bk\cdot \bde_1+ \cos \bk\cdot \bde_2+ \cos \bk\cdot \bde_3.
$
On the other hand,  $\psi(\bk)$ depends on the type of pairing under consideration. Note that $p$-wave pairing breaks lattice rotational symmetry. Therefore we consider $p+\I p$-wave and $f$-wave pairings:
$
  \psi^{f}(\bk)
  =
  \I(\sin \bk\cdot \bde_1- \sin \bk\cdot \bde_2+ \sin \bk\cdot \bde_3)$,
  $\psi^{pip}(\bk)
  =
  \I(
  \sin \bk\cdot \bde_1+ e^{\I\pi/3} \sin \bk\cdot \bde_2
  + e^{2\I\pi/3} \sin \bk\cdot \bde_3)
$.
Eq.~(\ref{Htriplet}) is solved with Bogoluybov transformation acting separately on each fermion species. This results in the spectrum
$
  E^\alpha_\bk
  =
  \sqrt{(\chi^\alpha_\bk/2)^2+|\D^\alpha_\bk|^2},
$
and mean field equations:
\begin{subequations}
\label{EMF-odd}
\bea
 \chi^{\alpha}
  &=&
  \frac{1}{N}\sum_\bk\frac16 \gamma(\bk) \left[1-\frac{\chi^\alpha_\bk}{2E^{\alpha}_\bk}\right] ,
  \\
  \D^{\alpha}_\dt
  &=&
  \frac{1}{N}\sum_\bk \frac13 \psi^*(\bk) \frac{\D^{\alpha}_\bk}{2E^{\alpha}_\bk}
  ,
  \\
  n^\alpha
  &=&
  \frac{1}{N}\sum_{\bk}
  \frac12\left[1-\frac{\chi^\alpha_\bk}{2E^{\alpha}_\bk}\right],
\eea
\end{subequations}
supplemented by the constraint equation $\corr{\vec{f}_\bi^\dagger \cdot \vec{f}_\bi} =1$.

\emph{Pairing in even channel.} Hoppings are defined as in~(\ref{chiD}), whereas pairing is
$
  \D^{xy}_\ds
  =
  1/2\corr{f_{\bi x}f_{\bi+\bde_1 y}-f_{\bi y}f_{\bi+\bde_1 x}}.
$
The Hamiltonian is:
\be \nn
  \tilde H
  =
  \sum_{\bk,\alpha}
  \chi^\alpha_\bk f^\dagger_{\bk \alpha}f_{\bk \alpha}
  +
  \D_\bk^{xy}  f^\dagger_{\bk x}f^\dagger_{-\bk y}
  +
  \D_\bk^{xy *}  f_{-\bk y}f_{\bk x},
\ee
with $\chi^\alpha_\bk$ given by Eq.~(\ref{ChiMF}), and
$ \label{Dsinglet}
  \D_\bk^{xy}
  =
  2 J \psi(\bk) \D^{xy}_\ds .
$
Note, that the \emph{$f_z$ band is unpaired and retains its Fermi surface.}
We consider $s$-wave and $d+\I d$-wave pairings ($d$-wave violates lattice symmetry and higher orbital momentum pairing requires inclusion of further neighbors). For the case of $s$-wave pairing, the function $\psi^{s}(\bk)= \gamma(\bk)$. For $d+\I d$-wave pairing we have $\psi^{did}(\bk)=\cos\bk\cdot\bde_1 + e^{2 \I \pi/3} \cos \bk\cdot\bde_2+e^{-2 \I \pi /3} \cos  \bk\cdot\bde_3$. The Bogolyubov  spectrum is
$  E^{x}_\bk
  =
  E^{y}_\bk
  =
  \sqrt{(\chi^{x,y}_\bk)^2+|\D^{xy}_\bk|^2}$,
$
  E^z_\bk
  =
  \chi^z_\bk.
$
Self-consistent mean field equations for $x$ and $y$-components are given by Eq.~(\ref{EMF-odd}) with the new expressions for the spectrum and gap functions. For the $z$ component we have
\be \nn
  \chi^z
  =
  \frac{1}{N}\sum_\bk\frac13\gamma(\bk) n_F(\chi^z_\bk)
  ,
  \qquad
  n^{z}
  =
  \frac{1}{N}\sum_\bk n_F(\chi^z_\bk)
  .
\ee

Our mean field approach includes on-site FN order  automatically. The on-site nematic order is described by the order parameter tensor, $Q^{\alpha\beta}=1/2\corr{\bS^\alpha\bS^\beta+\bS^\beta\bS^\alpha}-2/3\delta^{\alpha\beta}$. For a single site with $S=1$ all states with zero average spin $\corr{\bS}=0$ can be characterized by the unit director vector $\bn$~\cite{Lauchli}, in the basis defined earlier, $|\bn\rangle= \n_x |x\rangle + \n_y|y\rangle + \n_z|z\rangle$. For this state $Q^{\alpha\beta}$ is expressed via $\bn$ as $Q^{\alpha\beta}=1/3\delta_{\alpha\beta}-\n_\alpha \n_\beta$. For example,  $\bn\| \hat z$ corresponds to the state $|S^z=0\rangle$, and the nematic order is diagonal,
$Q^{\alpha\beta}=\mathrm{diag}(1/3,1/3,-2/3)$. In our model we also have states with vanishing spin order and diagonal on-site nematic order. However, since our GS is RVB-like with long-range entanglement, $Q^{\alpha\beta}$ cannot be described by the above simple form. We have to introduce the magnitude $q$,
$Q^{\alpha\beta}=q(1/3 \delta_{\alpha\beta}-\n_\alpha \n_\beta)$. Calculating the nematic order parameter tensor in our model we have $Q_\bi^{\alpha\beta}=\delta_{\alpha\beta}[1/3-n^{\alpha}]$, where $n^{\alpha}$ is the average occupation of corresponding fermion. Since $n^{x}=n^{y}$, we have nematic order with $\bn\| \hat z$, with a magnitude given by $q=n^{z}-n^{x}$, varying from $1$ for $n^z=1$ (state $|S^z=0\rangle$) to $-1/2$ for $n^z=0$. Non-zero anisotropy $D\neq0$ causes $n^{\alpha}$ to be different from $1/3$, and therefore directly couples to FN order along the $z$-axis.

\begin{figure}
\includegraphics[width=.9\linewidth]{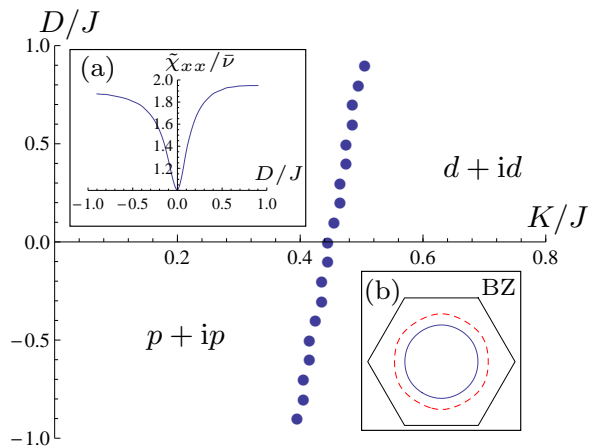}\\
\caption{\label{Fig:Plane}
The phase boundary between SL GS's with $p+\I p$ and $d+\I d$ pairing. (a) The spin susceptibility  $\tilde \chi_{xx}$ in the $d+\I d$ phase as a function of $D/J$ for $K/J=0.55$. The susceptibility is normalized by the average density of states, $\bar \nu= (\nu_x+\nu_z)/2$, where  $\nu_x$ is calculated without the gap. (b) Gapped (dashed red line) and ungapped (blue line) Fermi surfaces of  $x$, $y$, and $z$-fermions for $K/J=0.55$, $D/J=0.8$.
}%
\end{figure}%

Having studied the energies of all aforementioned states using Eq.~(\ref{Egsp}), we found that the main competition is  between states with $p+\I p$ and $d+\I d$-wave pairings, all other states being higher in energy. As one increases $K$, the effective coupling for the odd-channel pairing decreases, whereas for even pairing it remains the same. Finally for $K\approx 0.45 J$ singlet pairing wins. The resulting phase diagram is shown in Fig.~\ref{Fig:Plane}. The boundary between two states appears to be weakly dependent on~$D$.

\emph{Physical properties of $d+\I d$ state.} The $d+\I d$ state breaks time-reversal symmetry. The chiral order parameter associated with this broken symmetry $\corr{\bS_{\bi}\cdot(\bS_{\bi+\bde_1}\times \bS_{\bi+\bde_2})}\propto \chi^z |\D^{xy}_\ds|^2$  is proportional to the magnitude of the pairing gap squared. In addition, pairing with $d+\I d$ gap symmetry in two dimensions is topological~\cite{Senthil-did}, resulting in the existence of a pair of zero-energy edge modes at the boundaries. The physics of these modes will be discussed elsewhere.

The combination of gapless excitations with topological pairing gives rise to a number of unusual physical properties, that may explain the results of the recent experiment~\cite{exp}. Due to ungapped $f_z$  excitations the specific heat depends linearly on temperature near $T=0$, $C=\pi^2 k_B^2 \nu_z T/3 $, where $\nu_z$ is the density of states of $f_{\bi z}$ at the Fermi surface. Due to Higgs mechanism the gauge field is massive and does not modify the linear in $T$ behavior of the specific heat. The spin susceptibility exhibits more exotic behavior: due to the pairing of $x$ and $y$-fermions the $zz$-component $\chi_{zz}=0$. On the other hand, $\chi_{xx}$ is finite and depends on the anisotropy $D$. For $D$ smaller than the gap, $\tilde \chi_{xx}=\chi_{xx}/(\mu_B g)^2\approx \nu_z$, and approaches a factor two larger value  $\tilde\chi_{xx}\approx 2\nu_z$, when $D$ is much larger than the gap. This difference by factor $2$ is approximate, valid in the limit of constant gap and density of states. The behavior of $\tilde \chi_{xx}$ is shown in Fig.~\ref{Fig:Plane}~(a). We calculate Wilson ratio defined as $R_W = (4\pi^2 k_B^2)/(3 g^2\mu_B^2) (\bar \chi T)/C$, and obtain $R_W=8/3\approx 2.66$ for the case of small anisotropy, and $R_W \rightarrow 16/3\approx 5.33$ for large anisotropy. Note that we take the average susceptibility $\bar \chi=2/3 \chi_{xx}$ to account for the polycrystalline nature of the sample. The latter value gives surprisingly good agreement with the Wilson ratio observed experimentally, $R_W\approx 5.63$. We also calculated the imaginary part of the spin susceptibility. Since two out of three fermions are gapped, $\Im \chi_{\alpha\alpha}(\omega,\mathbf q)$ vanishes for temperatures and frequencies smaller than the gap for all $\alpha$. This implies the NMR relaxation $1/(T_1 T)$ is exponentially small for temperatures below the pairing scale. These results tell us that the Fermi surface associated with $f_z$~[see Fig.~\ref{Fig:Plane}~(b)] should be viewed very differently than the spinon Fermi surface in the $S=1/2$ SL which carries spin-$1/2$ quantum numbers and leads to gapless spin-$1$ excitations. In our case $S^z=1$ excitations are gapped even though the static spin susceptibility $\chi_{xx},\chi_{yy}\neq0$ and the specific heat has linear $T$ dependence.

Finally, we discuss experiments that could confirm the proposed ground state. Measurement of the spin susceptibility for single crystal or oriented powder samples is of great interest in order to test our prediction of strong anisotropy. We also predict an exponentially activated behavior for $1/(T_1 T)$ which may be surprising in view of the linear $T$ behavior of the specific heat.

We thank Luis Balicas for bringing Ref.~\cite{exp} to our attention. We acknowledge useful discussions with Samuel Bieri. T.S. is supported by grant {NSF-DMR}~{6922955}. P.A.L. is supported by {NSF-DMR}~{1104498}.
%
\end{document}